# Integration of hydrothermal liquefaction and carbon capture and storage for the production of advanced liquid biofuels with negative $CO_2$ emissions

E.M. Lozano, T.H. Pedersen[*], L.A. Rosendahl

Department of Energy Technology, Aalborg University, Pontoppidanstræde 111, 9220 Aalborg Øst, Denmark

**Abstract:** The technical and economic feasibility to deliver sustainable liquid biocrude through hydrothermal liquefaction (HTL) while enabling negative carbon dioxide emissions is evaluated in this paper, looking into the potential of the process in the context of negative emission technologies (NETs) for climate change mitigation. In the HTL process, a gas phase consisting mainly of carbon dioxide is obtained as a side product driving a potential for the implementation of carbon capture and storage in the process (BECCS) that has not been explored yet in the existing literature and is undertaken in this study. To this end, the process is divided in a "standard" HTL base and a carbon capture add-on, having forestry residues as feedstock. The Selexol™ technology is adapted in a novel scheme to simultaneously separate the $CO_2$ from the HTL gas and recover the excess hydrogen for biocrude upgrading. The cost evaluation indicates that the additional cost of the carbon capture can be compensated by revenues from the excess process heat and the European carbon allowance market. The impact in the MFSP of the HTL base case ranges from -7% to 3%, with -15% in the most favorable scenario, with a GHG emissions reduction potential of 102-113% compared to the fossil baseline. These results show that the implementation of CCS in the HTL process is a promising alternative from technical, economic and environmental perspective in future scenarios in which advanced liquid biofuels and NETs are expected to play a role in the decarbonization of the energy system.

**Graphical abstract:**

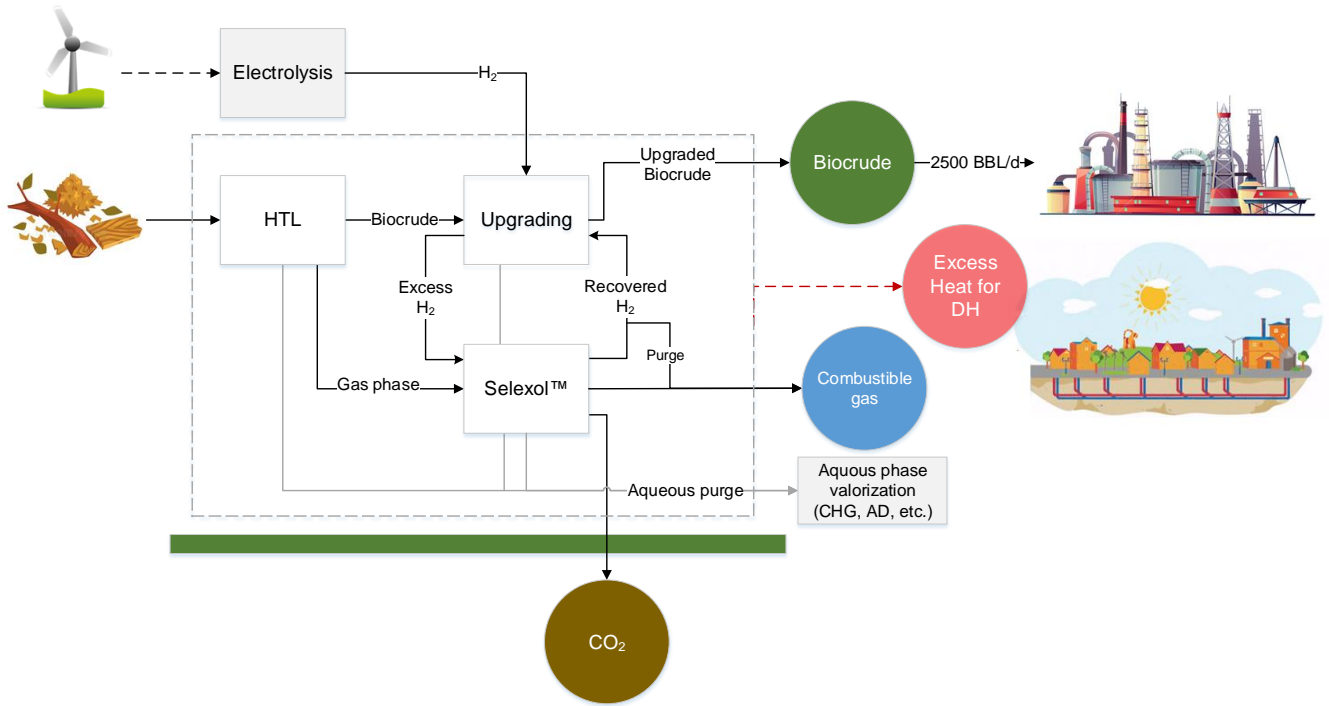

## 1. INTRODUCTION

According to the IPCC's fifth assessment report, negative emission technologies (NET's) are likely to play a significant role to meet the climate targets in the future and hold global warming to well below 2 ºC compared to pre-industrial levels, as established in the Paris agreement. Among these, the combined use of biomass with carbon capture and storage (BECCS) has shown the highest $CO_2$ reduction potential in Integrated Assessment Models (IAMs), being identified as a key concept to achieve climate change mitigation [1,2].

[*] Corresponding author. E-mail address: thp@et.aau.dk

In BECCS, the combination of CCS with low-carbon or carbon-neutral bioenergy has shown capability to generate negative $CO_2$ emissions while simultaneously supply energy in the form of electricity, heat or fuel. In this approach, part of the carbon dioxide that is naturally absorbed by sustainably harvested biomass is not released back to the atmosphere but instead sequestered underground, having as result overall negative carbon emissions.

While the main focus of BECCS has been given to ethanol fermentation processes and cogeneration, as they comprise the main commercially available applications for extended biomass use, there is lack of literature that evaluates the implementation of CCS in thermochemical processes, and particularly in the production of advanced liquid biofuels, despite of its high potential and the increasing interest in the development and commercialization of these technologies.

Due to their compatibility with existing fossil-fuels infrastructure and utilization pathways, advanced drop-in biofuels are considered the most readily available alternative for direct implementation in the energy system, enabling a faster carbon emission reduction in the segments of the transportation sector that cannot be easily decarbonized by e.g. direct electrification (i.e. aviation, maritime). In Europe, the revised Renewable Energy Directive (REDII) entered into force in December 2018 setting a target of 14% renewables in the transport sector by 2030, with a minimum contribution of 3.5% for advanced biofuels [3].

Hydrothermal Liquefaction (HTL):

Among different technologies for advanced biofuels production, hydrothermal liquefaction (HTL) stands out as a highly feedstock flexible and conversion effective route, being capable of handling a wide range of wet/dry organic materials. HTL is a thermochemical process in which the main product obtained is a biocrude – an oxygenated precursor of hydrocarbon fuels – that can be further upgraded to meet product specifications using standard refinery technology. The HTL process occurs in the presence of water at sub or supercritical hydrothermal conditions, effectively in the range of temperatures from 250 ºC to 450 ºC, and pressures from approximately 100–350 bar for sufficient time to break down the polymeric structure of the biomass and to form the oily product (order of minutes). In a subsequent hydrotreating step, excess hydrogen is used for heteroatom removal, bonds saturation and overall improvement of the biocrude quality towards the final drop-in fuels [4].

HTL technology is currently at pilot/demonstration scale with several companies and on-going projects aiming to bring the technology to commercialization on different biomass types and at different process conditions. Examples are Steeper Energy ApS (Canada/Denmark), Licella (Australia), Muradel Pty Ltd. (Australia), Southern Oil Refining (Australia), Genifuel Corporation (USA), and Reliance (India). Recent estimates have shown that the minimum fuel selling price (MFSP) of finished fuel (e.g., gasoline equivalents) achieved using HTL technology is in the range of 0.6-1.3 USD/L, with the hydrogen consumption and price as one of the most sensitive parameters [4–6].

In the HTL literature, different feedstock have been tested including lignocellulosic biomass, algae and waste biomasses such as manure, sewage sludge or municipal solid waste. From these, lignocellulosic feedstock, particularly residual woody biomasses, represent the largest availability of non-food/low indirect-land use change (ILUC) biomass in Europe for advanced biofuels production due to the large volumes produced mainly as forestry and agriculture residues, more notably in the Nordic region [7]. Based on this and due to the availability of data at pilot scale, forestry residues is selected as feedstock for the study carried out in this paper in order to ensure consistency and relevance.

Alongside fuel production, a gas phase side product stream consisting mainly of carbon dioxide drives a potential for the implementation of carbon capture in the process that is of main interest in this study.

Selexol™ technology for carbon capture:

The process to capture $CO_2$ from a gaseous stream has been thoroughly studied in literature through pre-combustion, oxy-fuel and post-combustion methods, being gas-liquid absorption one of the most common and commercially mature technologies.

In general, for post-combustion capture the pressure of the gas is close to atmospheric and the $CO_2$ concentrations are relatively low (4-15 vol.% approximately), therefore requiring large amounts of solvent and energy for its regeneration and reutilization in the process. In cases where the concentration of $CO_2$ and pressure are higher, which is the case for the HTL gas, absorption with physical solvents is recommended. Typically, the $CO_2$ concentration is 25-40 vol.%, the pressure is normally between 20-70 bar and temperature around 30-40 ºC [8]. Among the main processes based on physical solvents, Selexol™, Rectisol® and Purisol®, the Selexol™ process has been one of the most studied for acid gas removal from high pressure syngas, where it has shown lower energy consumption with recoveries and purities of the captured $CO_2$ around 96-99 % [8–10]. Based on the technical suitability, and being a commercially mature technology, the Selexol™ process is selected to evaluate the carbon capture from the HTL gas.

Another relevant aspect to consider in the HTL process is the hydrogen supply for the biocrude upgrading. Likewise hydrotreating of fossil crudes in conventional refinery operations, hydrogen availability and the recovery of the excess hydrogen used is a critical issue for the economy of the HTL process. Physical absorption has been used industrially for hydrogen purification, having the advantage that the purified hydrogen is near the feed pressure. For other methods such as pressure swing adsorption (PSA) and membranes, higher purities can be achieved at expense of higher pressure drops and lower recoveries [11], for which absorption has been identified as an economic alternative for hydrogen recovery in refinery operations when the feed pressure is high [12]. Therefore, in standalone operation, the implementation of the Selexol™ technology brings the novel opportunity to study a dual function scheme for hydrogen recovery along with the carbon capture feature that has a capital cost reduction potential.

Overall, the literature review carried out shows that, to date, there is lack of studies that delve into the different technical and economic aspects of the HTL + carbon capture integration. Interestingly, in a recent publication, Cheng, Porter and Colosi evaluated the performance of hydrothermal treatment technologies (HTT) –namely hydrothermal carbonization, hydrothermal liquefaction and hydrothermal gasification- with CCS as NET for a different type of feedstock an process conditions [13]. In this study, the yields and characteristics of the products from HTT where estimated using machine learning tools, and $CO_2$ was assumed to be captured via an amine-based system with a fixed efficiency of 85%. The authors concluded that best overall energy and global warming reduction performance was achieved for HTT of lignocellulosic biomass at low temperature, however, supercritical conditions are not evaluated and the yields are relatively low, as they are mainly estimated based on reported data in batch operation. Furthermore, the economic evaluation was not part of the study, but an interesting comparison between conventional BECCS –direct combustion of biomass- and HTT+CCS is presented.

Given the research gaps encountered, the purpose of this paper is to evaluate the novel integration of the Selexol™ technology in the HTL process from a technical, economical and sustainability perspective having forestry residues as feedstock, based on published data for continuous operation at pilot scale.

The structure of the paper comprises a methodology section describing the modeling approach, followed by results and discussion and finally conclusions. In the first section, the methodology applied and modeling assumptions/considerations are presented for the standard HTL base case with upgrading and the carbon capture via Selexol™. Next, the results are discussed in terms of: 1) mass and energy balances of the overall process; 2) process design and sensitivity of the Selexol™ process, 3) heat integration to assess potential excess heat production, 4) cost estimation based on previous techno-economic analysis on HTL, and 5) GHG emission assessment. The price of the captured $CO_2$ and its impact in the MFSP of the HTL base case is established. The estimated GHG emissions of the process are compared to the fossil baseline, in order to assess the environmental impact and emissions reduction potential of the HTL+Selexol™ integration.

## 2. PROCESS DESCRIPTION AND METHODOLOGY

The process is designed to produce drop-in biofuels from forestry residues with an output of about 2,500 barrels per day as baseline. This is set in perspective of the RED II target and corresponds to 3.5 % of the capacity of a medium-size refinery of 68,000 barrels per day (Shell refinery in Fredericia, Denmark). Fractionation of the product to obtain the final fuels is not included and the upgraded oil is meant to be integrated to a conventional refinery for final processing. The overall model consists of 3 major process sections shown in Figure 1, -HTL, upgrading (UPGR) and Selexol-  each of which consists of multiple unit operation models (not shown in the high-level flowsheet diagram in Figure 1).

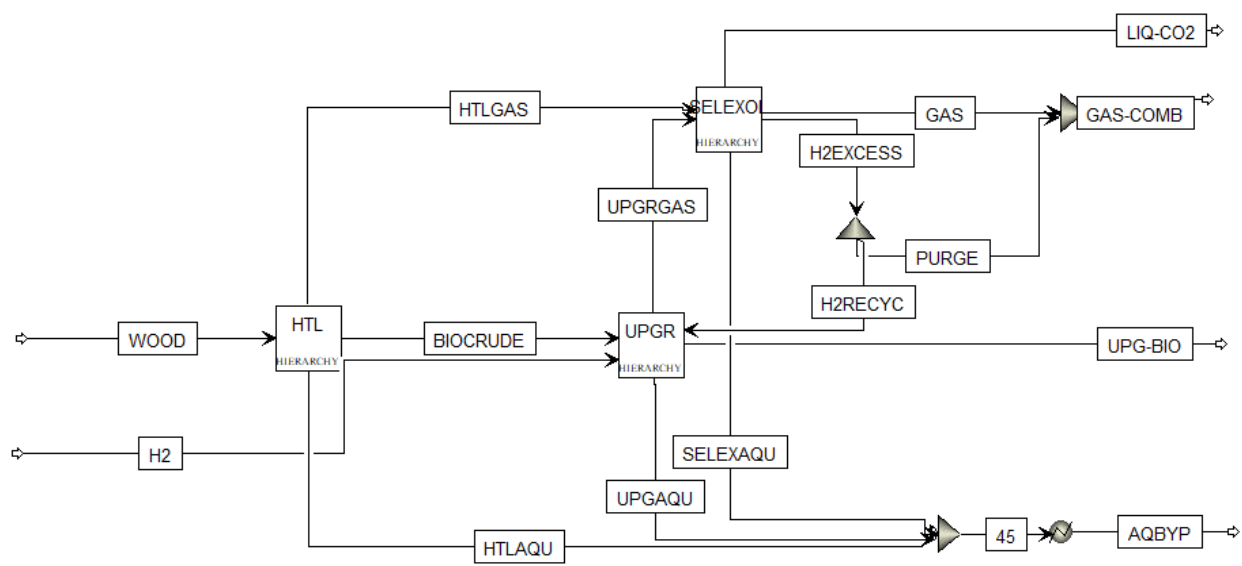

*Figure 1 High level HTL base + carbon capture process flowsheet in Aspen Plus®*

The process is modeled in Aspen Plus V9 using the Soave-Redlich Kwong (SRK) property package and the HTL and upgrading sections are built based on experimental results published in literature for woody biomass [14][15]. The Selexol™ process is modeled independently in Aspen Hysys V9 using the PC-SAFT property package, recommended by Aspen Tech with available binary parameters for modeling the absorption of $CO_2$ using the physical solvent DEPG. Due to the novelty of the HTL+ carbon capture process, there is not experimental data on the carbon capture of the HTL gas that can be used to contrast the modeling results, however, there is published documentation by Aspen Tech® that validates the PC-SAFT property package for carbon dioxide absorption with experimental and plant data at different process conditions [16]. The results from the Hysys model are incorporated into the Aspen Plus simulation for the overall mass and energy balances of the HTL+ carbon capture scheme. Minimum utility requirements are estimated using pinch analysis with $\Delta T_{min}$ of 20 ºC (typically 10-20 ºC [17]). A more detailed description of the hierarchy levels is presented in sections 2.1 and 2.2, and the methodology for the cost estimation and GHG emission assessment is presented in sections 2.3 and 2.4 respectively.

### 2.1 Standard HTL base with upgrading

This section comprises the production of biocrude through HTL with downstream hydrogenation in order to make it suitable for refinery processing. Table 1 shows the elemental composition of the biomass, HTL biocrude and hydrotreated biocrude used as reference in this study. The inputs of the main units in the hierarchy levels HTL and UPGR are summarized in Table 2. Other units such as pumps and heat exchangers are required to adjust to the process conditions indicated, but are not presented in Table 2.

*Table 1 Elemental composition of woody biomass, HTL biocrude and hydrotreated biocrude*

|  | C [wt.%] (daf) | H [wt.%] (daf) | O [wt.%] (daf) | HHV [MJ/kg] (daf) |
|---|---|---|---|---|
| Woody biomass | 49.1 | 5.9 | 43.7 | 19.9 |
| HTL biocrude | 80.0 | 8.4 | 11.0 | 35.9 |
| Hydrotreated biocrude | 87.4 | 12.6 | 0.0 | 43.9 |

*Table 2 Main input parameters of HTL and UPGR hierarchy levels in Aspen Plus®*

| Hierarchy | Unit | Model | Description | Block inputs* |
|---|---|---|---|---|
| HTL | HTL reactor | User2 model linked to Excel file. | • T=400 ºC, P= 300 bar.<br>• Biomass(dry-ash free):aqueous phase=1:3.2 | • Gas yield, $Y_{g/w}$=0.412 kg/kg [14]<br>• Biocrude yield, $Y_{bc/w}$=0.453 kg/kg [14]<br>• Gas composition (Table 3)<br>• Biocrude composition [18] |

| | Separator | Component splitter | • TOC=70-75 g/L [14]. <br> • T= 150 ºC, P= 40 bar | • Split fractions |
|---|---|---|---|---|
| | Evaporator | Evaporator | • P= 3 bar | • Design specification to fix flow of recycled aqueous phase according to biomass to aqueous phase ratio, varying cold stream outlet vapor fraction |
| UPGR | Hydrotreater | User2 model linked to Excel file. | • T=370 ºC, P= 66 bar <br> • Make up hydrogen adjusted by design specification. | • Gas yield, $Y_{g/bc}$=0.05 kg/kg [19] <br> • Biocrude yield, $Y_{bc/bc}$=0.85 kg/kg [19] <br> • Gas composition (Table 3) <br> • Biocrude composition (pseudo-components). <br> • C/H/O %wt. of upgraded biocrude [19] |

*Subindices g= gas, bc= biocrude, w=woody biomass.

### 2.1.1 Hydrothermal Liquefaction modeling of woody biomass

A slurry of biomass and water is fed to the HTL reactor and the products are cooled down and decompressed to be separated. The biomass is defined as a non-conventional solid and the default settings of HCOALGEN are modified for a more accurate representation of the heat of formation and heat capacity as described by Lozano et al. in [18]. The biocrude is modeled using the model compounds approach, adjusting its composition by multi-objective optimization to meet the physical and thermal properties reported, as discussed in detail in a previous publication [18]. The composition of the gas phase is known (Table 3) and the composition of the aqueous phase is adjusted based on reported TOC during aqueous phase recirculation [15]. The mass balance is set according to the reported yields of the Hydrofaction® process [14] presented in Table 2, in which solids or char production was not observed experimentally in significant amounts in continuous operation. The mass balance of each component (i) in each produced phase (j) is estimated as:

$$m_i^j = m_i^{feed} + (m_w Y_{j/w}) z_i^j \qquad (1)$$

Where $m_i^{feed}$ is the mass flow of the component (i) in the feed, $m_w$ is the mass flow of woody biomass, $Y_{j/w}$ is the yield of the produced phase (j) from Table 2, and $z_i^j$ is the know composition of the compound (i) in phase (j). After the reactor, cooling and expansion is necessary for the separation, and the conditions are chosen to match the typical pressure range of the Selexol™ process (40-70 bar), taking into account the separation system described for the HTL pilot plant [15]. Since water is produced in the reaction, it is assumed that the excess fraction of the aqueous phase is separated by evaporation and the remaining is recirculated for the slurry preparation. The yield of the total aqueous phase is calculated by difference from the reported yields of gas and biocrude.

### 2.1.2 Hydrotreater modeling of biocrude

The HTL biocrude is fed to the hydrotreating unit with hydrogen in excess and the phases produced are calculated following a similar procedure described above based on yields and composition from experimental results. In this case, the gas and upgraded biocrude yields are known, as well as the gas composition. The upgraded biocrude is modeled with the petro-characterization tools available in Aspen Plus®, and not with the model compounds approach used for the HTL biocrude. Experimentally obtained distillation curve and specific gravity of the hydrotreated biocrude are used to perform a pseudo-components breakdown for further property estimations [15].

The difference in the oxygen content between the initial and the upgraded biocrude is used to estimate water production via hydrodeoxygenation. Since there is no data available on the composition of the aqueous phase produced through hydrotreatment, one component was selected as representative of the organic loading of this phase in order to close the mass balance and minimize atom imbalance across the process.

The hydrogen consumption is fixed at 0.04 g/g oil, which is in the range reported in [19], and the total available hydrogen is set as 10 times the required (0.4 g/g oil). The production of other gas phase compounds is determined based on the composition of the gas effluent reported in Table 3 for a larger excess (hydrogen consumption of 1.9% of the initial). The excess hydrogen is further purified in the Selexol™ process and recirculated to the reactor. Due to the presence of impurities –mainly methane- a purge is necessary before recirculation and a make-up of pure hydrogen is required to compensate for the consumption and the losses.

Typically, the purge can be established targeting a threshold value of the methane, however, this analysis requires the knowledge of how methane changes across the hydrotreater. Consequently, a purge ratio of 0.3 (volume of hydrogen in the purged gas/volume of hydrogen in the make-up gas), equivalent to approximately 5 % of the mass flow, was selected based on typical purge requirement of hydrodesulfurization (HDS) processes reported for petroleum refining, which are in the range of 0.1 (Naptha) to 0.3 (VGO) [20]. Due to limitations in the experimental data, it is assumed that the impurity in the recycle does not affect the products characteristics or the hydrotreating efficacy. The make-up is calculated by design specifications to ensure a fixed hydrogen inlet in the reactor corresponding to the hydrogen availability of 0.4 g/g oil.

*Table 3 Composition of HTL and hydrotreater effluent gases and relative solubility in Selexol™ solvent*

|  | HTL gas Average dry [vol. %] [14] | Hydrotreater gas [vol. %] [19] | Relative solubility in Selexol™ at 25 ºC ($CO_2$=1) [16,21] |
|---|---|---|---|
| $H_2$ | 25.80 | 93.9 | 0.01 |
| $CO_2$ | 61.10 | 1.30 | 1.00 |
| CO | 0.30 | 0.90 | 0.03 |
| Methane | 7.20 | 2.30 | 0.07 |
| Ethene | 0.20 | 0.00 | -- |
| Ethane | 2.40 | 0.90 | 0.42 |
| Propene | 0.30 | 0.00 | -- |
| Propane | 1.00 | 0.50 | 1.01 |
| Butane | 0.70 | 1.10 | 2.37 |
| Methanol | 0.40 | 0.00 | -- |
| Ethanol | 0.30 | 0.00 | 257 |
| Acetone | 0.30 | 0.00 | -- |
| $H_2O$ | 0.00 | 0.00 | 730 |
| Total | 100.00 | 100.00 |  |
| HHV [MJ/kg] | 7.73 | 81.67* |  |

*Calculated assuming ambient conditions (P=1 atm, T=25 ºC)

### 2.2 Carbon dioxide capture and hydrogen purification- Selexol™

This section comprises the Selexol™ process adapted to serve two functions simultaneously: 1) to separate the carbon dioxide from the HTL gas for subsequent cooling and compression, and 2) to purify the hydrogen from the hydrotreater gas effluent that needs to be recycled for the economy of the process.

The solvent used in this process is a mixture of dimethyl esters of polyethylene glycol (DEPG) with high solubility for the acid gases $CO_2$ and $H_2S$ relative to $H_2$, CO and methane. Other light hydrocarbons such as ethane and propane have similarly high solubilities, while higher hydrocarbons (C3+), alcohols and water are significantly more soluble in the solvent and can decrease the final purity of the $CO_2$, required to be above 95 vol.% to avoid negative impacts on the pipelines and disposal sites [22]. The purification limits for transportation of $CO_2$ are shown in Table 4. An average composition of the HTL gas is used as shown in Figure 2 based on data reported in literature for lignocellulosic feedstock [23,24].

*Table 4 Purification limits for pipeline transportation of $CO_2$ stream from CCS [22]*

| Component | Overall range for requirement levels |
|---|---|
| $CO_2$ | >95 vol.% |
| CO | <2000 ppmv |
| Hydrocarbons (HCs) | <5% vol.% |
| $H_2O$ | <50 ppmv |
| $O_2$ | <10 ppmv |
| $N_2$/$H_2$/Ar (All non-condensables) | <4 vol.% |

Different configurations are available in literature for the Selexol™ process that depend mainly on the characteristics of the gas. The process here designed (Figure 3) is based on existing layouts for syngas [8,9] and was adjusted for the dual functionality of the process. It consists of two absorption columns for the two different gas inlets –one for the effluent gas from HTL and one for the hydrotreater effluent gas- which are fed at different pressures. The regenerated solvent is first utilized in the higher pressure absorber (60 bar) to remove hydrocarbon impurities from the excess hydrogen, therefore hydrogen at higher purity is obtained.  The solvent with relatively low load of impurities is subsequently fed to the 40 bar pressure absorber, where the $CO_2$ is absorbed and separated from the combustible gases in the HTL effluent that have low solubility in the solvent.

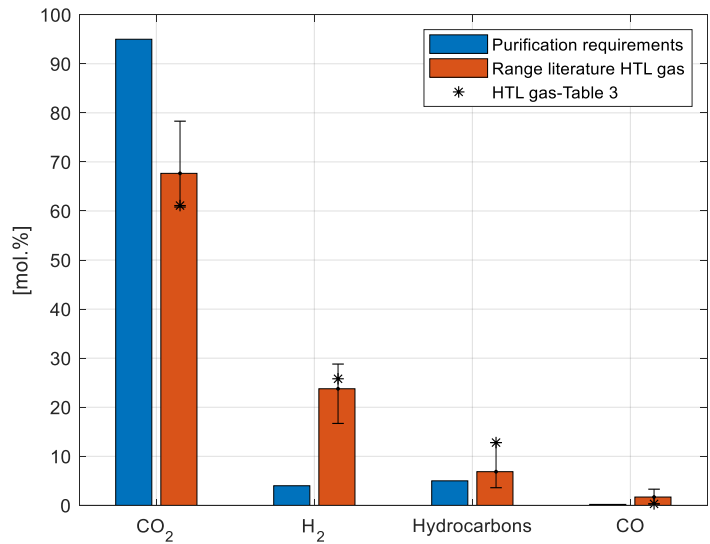

*Figure 2 Composition of HTL gas compared to purification limits for pipeline transportation*

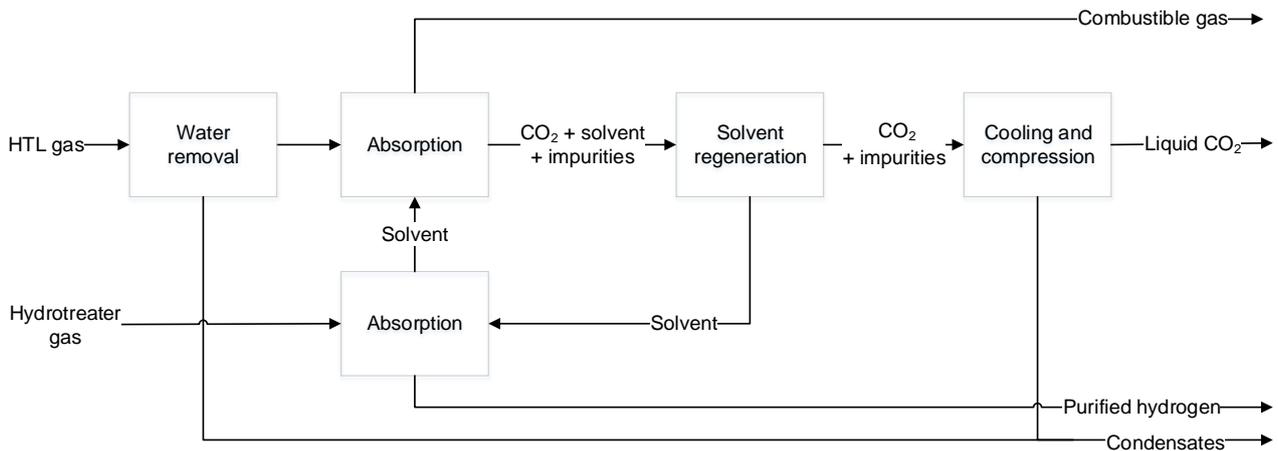

*Figure 3 Simplified scheme of Selexol™ process with dual functionality modeled in Aspen Hysys*

A more detailed description of the main units is presented as follows: The HTL gas is fed at 150 ºC and 40 bar. In order to account for the presence of water, the gas in dry basis (Table 3) is first saturated with water vapor (relative humidity of 100 %) at the input conditions and is subsequently cooled down in order to separate condensable gases before entering the absorption column. The removal of condensables –mainly butane, alcohols and water- before the absorption is necessary due to their high solubility in the DEPG solvent. The presence of these components in the absorption column increases the solvent requirement and affects the regeneration process; furthermore, valuable hydrocarbons that cannot be recovered from the solvent constitute a loss for the heating potential of the clean gas. This is considered not necessary for the hydrotreater gas as the reaction does not take place in presence of excess water.

In the absorption column the solvent is fed at the top and is contacted with the gas in a countercurrent operation. The temperature, number of stages and solvent flow are the parameters analyzed to favor high selectivity of $CO_2$. The gas produced from the top is analyzed for heat supply in the process and potential excess heat production for district heating use. The enriched solvent is regenerated by means of successive expansion and separation steps to release the $CO_2$ and impurities. A reboiled column is used to remove the compounds that are more strongly absorbed, increasing the purity of the solvent before the absorption of the hydroteater gas.

In the second absorber, the effluent gas from the hydrotreater is fed at 60 bar and relatively low temperature of 25 ºC, in order to discard the presence of water and other condensates. The $CO_2$ is delivered for cooling and compression and the hydrogen is recycled to the hydrotreating unit.

The impact of different parameters is evaluated by means of sensitivity analyses targeting high purity and recovery of the obtained $CO_2$ and purified hydrogen. The sensitivity with changes in the composition of the HTL gas is also analyzed, varying the $CO_2$ and summed hydrocarbons content within the ranges shown in Figure 2, keeping the ratios of the other components constant.

### 2.3 Cost estimation

The minimum fuel selling price (MFSP) is estimated by means of net present value analysis (NVP), and it corresponds to the price at which the produced fuel should be sold to achieve a zero equity net present value for a project life of 25 years and a discount rate of 10 %. The total capital investment takes place at the beginning of the project and the annualized operational costs are estimated based on the simulation results, establishing the utilities consumption after heat integration. The prices used for raw materials, utilities and other cost components are reported in Table 5.

For the HTL baseline with upgrading, an estimation of the total capital investment is reported in previous studies for a plant of similar size (1,000 ton per day of organic matter) [6,25], and is used in this study for the calculation of the MFSP without the Selexol™ add on.

For the Selexol™ process, the total capital investment is estimated with the Aspen Process Economic Analyzer tool available in Hysys, and the result is compared to the reported in literature of carbon capture facilities in order to assess the reliability of this estimation. The estimation is performed based on the simulation results following a three-step procedure: mapping of unit operations to process equipment, sizing and evaluation of equipment cost. Regarding the solvent price, it was not possible to find reliable standard market prices as the solvent is not traded as a commodity. However, the price used is comparable to the reported for other solvents such as MEA [26]. The solvent cost is added to the capital cost, as it is not estimated in Hysys, and it is assumed that replacement of solvent is not necessary in the project lifetime, as it has been reported that DEPG is a stable solvent and is very resistant to degradation, even with oxygen [27] [28]. Total operational cost of the Selexol™ process are estimated from the simulation results.

The cost of the avoided $CO_2$ is calculated by means of NVP analysis based on the capital and operational costs of the Selexol™ process, likewise as for the MFSP with the same project lifetime and discount rate. The result is compared with the range reported for other BECCS applications available in literature (20-175 USD/ton $CO_2$) [2], and the ratio of $CO_2$ to fuel produced is used to estimate the impact of the carbon capture process in the MFSP of the baseline.

Subsequently, the impact of revenues from excess heat sale and $CO_2$ European emission allowances are evaluated. In the case of the heat revenues, a fixed price was maintained for the minimum and maximum cases but it was assumed that just 30%, 40% and 50% of the heat was available in the minimum, base and maximum scenarios in order to have a more realistic view on efficiency losses.

Finally, Monte Carlo simulations were performed with a random variation of the cost parameters in Table 5 in order to estimate the MFSP range expected when the cost of the Selexol™ process and the revenues from heat and the $CO_2$ European emission allowances are accounted.

*Table 5 Parameters for cost estimation*

|   | Cost component | Base | Min | Max | Reference |
|---|---|---|---|---|---|
| CAPEX | Total Capital Investment of HTL + UPGR plant [million USD] | 225.80 | 119.30 | 305.30 | [6] |
| Variable Operational costs (VOC) | Wood residue [USD/tonne] | 41.50 | 37.00 | 70.00 | [6] |
|  | Hydrogen [USD/kg] | 5.45 | 3.11 | 6.28 | [29] |
|  | DEPG [USD/kg] | 3.50 | 1.50 | 5.50 | [26] |
|  | Wood grinding energy [MWh/tonne] | 0.13 | 0.02 | 0.24 | [6] |
|  | Electricity [USD/MWh] | 76.3 | 43.6 | 88.0 | [30] |
|  | Fired heater [USD/MWh] | 15.30 | 10.00 | 21.80 | Aspen HYSYS |
|  | Cooling [USD/MWh] | 0.76 | 0.50 | 1.27 | Aspen HYSYS |
|  | Refrigerant [USD/MWh] | 9.76 | 8.78 | 10.73 | Aspen HYSYS |
|  | Water disposal | 2.5% of VOC | -- | -- | [6] |
| Fixed Operational costs (FOC) | Fixed Operational costs | 17.5% of VOC |  |  | [6] |
| Revenues | Excess heat [USD/GJ] | 11.08 | -- | -- | [31] |
|  | $CO_2$ European emission allowances [USD/tonne] | 26.16 | 19.62 | 31.61 | [32] |
|  | Annual operating hours | 8,000 |  |  |  |

### 2.4 GHG emissions assessment

The GHG emissions of the standard HTL with upgrading, with and without the carbon sequestration add-on, were evaluated in order to estimate the potential emissions reduction relative to the fossil-fuels baseline. The analysis includes combustion of the final product assuming diesel and heavy fuel usage in equal parts. The parameters used in the analysis are summarized in Table 6 and were mostly adopted from the reported in [19]. The emission factor used for the electricity corresponds to the average reported for the Danish system for the period 2018-2019 [33] and the emissions from hydrogen production via electrolysis are estimated based on an electricity consumption of 3.8 kWh/Nm$^3$ of hydrogen reported for commercial alkaline electrolyzers [34], which corresponds to an 85 % efficiency, approximately.

*Table 6 Emissions factors used in GHG emissions analysis*

| Source | Value |
|---|---|
| Feedstock supply [kg CO$_2$/tonne] | 44 |
| Emission factor electricity Denmark [kg CO$_2$/MWh] | 171.5 |
| Catalysts [kg CO$_2$/tonne produced fuels] | 39 |

### 3. RESULTS AND DISCUSSION

The overall mass balances and elemental balances of inputs and outputs in the high level model are shown in Figure 4 and Table 7. The detailed results summary of all the streams in the hierarchy blocks can be consulted in the Supplementary material. The overall hydrocarbon yield of the process is 38.32% relative to the biomass input, with a carbon efficiency of 68.2%. Carbon dioxide is the second largest product containing about 25% of the initial carbon. The errors in the elemental balances are below 5% and can be related to the balances in the HTL and hydrotreater discussed in the next sections.

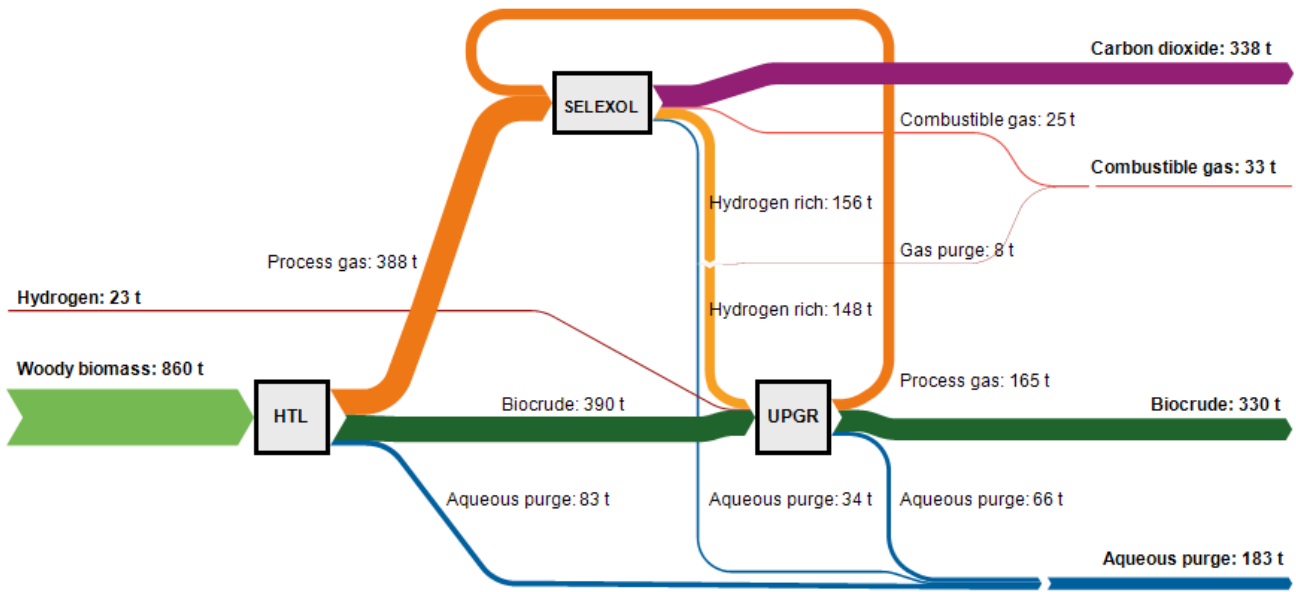

*Figure 4 Mass balance of the overall process* (aqueous phase recirculation of 2752 tonne/day in the HTL not included)

*Table 7 Simulation results of mass and elemental balance*

| [tonne/day] | Mass flow | | C | | H | | O | |
|---|---|---|---|---|---|---|---|---|
| | Feed | Product | Feed | Product | Feed | Product | Feed | Product |
| Woody biomass | 860.00 | 0.00 | 422.17 | 0.00 | 50.65 | 0.00 | 375.65 | 0.00 |
| H$_2$ | 23.34 | 0.00 | 0.00 | 0.00 | 23.34 | 0.00 | 0.00 | 0.00 |
| Biocrude | -- | 329.59 | -- | 288.06 | -- | 41.53 | -- | 0.00 |
| CO$_2$ | -- | 337.80 | -- | 98.12 | -- | 2.11 | -- | 237.57 |
| Aqueous phase | -- | 183.06 | -- | 34.74 | -- | 18.16 | -- | 130.16 |
| Gas for combustion | -- | 32.81 | -- | 11.10 | -- | 15.43 | -- | 6.28 |
| Total | 883.34 | 883.26 | 422.17 | 426.15 | 73.99 | 77.23 | 375.65 | 374.01 |
| Error (%) | -- | 0.01 | -- | 2.33 | -- | 4.38 | -- | 0.44 |

### 3.1 Standard HTL base with upgrading

Mass and elemental balance: The results of the elemental balances in the HTL reactor and hydrotreater are indicated in Table 8. A negative duty was obtained for both HTL and hydrotreater indicating net exothermic operation. Even though a positive duty was obtained for the HTL reactor in a previous study [18], in the present case the yields were modified and water production increased by 35%. This could explain the change in the energy balance due to the exothermic nature of this reaction.

*Table 8 Duty and elemental balances across HTL reactor and hydrotreater*

|  | HTL | UPGR |
|---|---|---|
| Duty [MW] | -6.38 | -5.94 |
| Error C (%) | 0.04 | 0.30 |
| Error H (%) | 3.06 | 0.32 |
| Error O (%) | 0.05 | 6.14 |

The errors obtained in the elemental balance across the reactors are expected due to the limitations in the model compound approach in the HTL biocrude and the limited knowledge on the composition of the aqueous phase in both HTL and upgrading reactions. The components used to represent the water soluble organics in HTL aqueous phase are phenol, glycolic acid, acetone and $C_1$-$C_2$ alcohols, in agreement with the reported by Madsen et al. [35] in the characterization of HTL products from woody feedstocks. The detailed composition of the streams can be consulted in the Supplementary material.

From the results, approximately 7% of the carbon in the aqueous effluent is transferred to the evaporated water and discarded in the aqueous purge. However, this value is expected to be lower due to the basic pH used in real conditions at which the deprotonated/ionic form of the species is dominant and therefore less volatile. In the hydrotreater, just one component –phenol- was used as the representative of the water soluble organics given the lack of experimental data. Overall, further adjustment of the composition of water soluble organics would be needed to improve the mismatch. The HTL and upgrading gas effluents are fed to the Selexol™ process.

Modeling of biocrudes: The results of the properties of the raw and upgraded biocrude are shown in Table 9 compared to experimental values reported, and the true boiling point (TBP) curves are presented in Figure 5. For the biocrude before upgrading the errors in the reported properties are minimized and the enthalpy of formation is accurately estimated aiming for a more reliable estimation of the heat of reaction, as discussed in [18]. The data shown in Figure 5 corresponds to the accumulated distribution of model compounds vs. their normal boiling points. It is observed that the distribution is such that it follows the trend of the distillation profile until 350 ºC, but it needs to be adjusted above this temperature for the heavy fraction to adequately represent the nature of the oil. Despite of this mismatch, the results in Table 9 show that the bulk thermal properties estimated are in good agreement with the reference values and are considered satisfactory for energy balance considerations, which is the main focus in the analysis presented. Since the HTL biocrude is not distilled or separated and the total produced is sent to the hydrotreater, a better fit of the TBP is not considered critical for the validity of the results at this stage.

*Table 9 Biocrude properties (1 bar, 25 ºC) compared to reference values*

|  | HTL | | UPGR | |
|---|---|---|---|---|
|  | Reference [15] | Simulation | Reference [15] | Simulation |
| HHV [MJ/kg] (daf) | 35.90 | 34.03 | 43.90 | 46.33 |
| Heat of formation [MJ/kg] | -2.24* | -2.24 | -2.62* | -2.07 |
| C [wt.%] | 80.00 | 77.16 | 87.40 | 85.35 |
| H [wt.%] | 8.40 | 10.36 | 12.60 | 14.64 |
| O [wt.%] | 11.00 | 12.48 | 0.00 | 0.00 |
| Specific heat [kJ/kg K] | NR** | 1.80 | NR** | 2.05 |

*Estimated from experimental HHV. **NR: Not reported

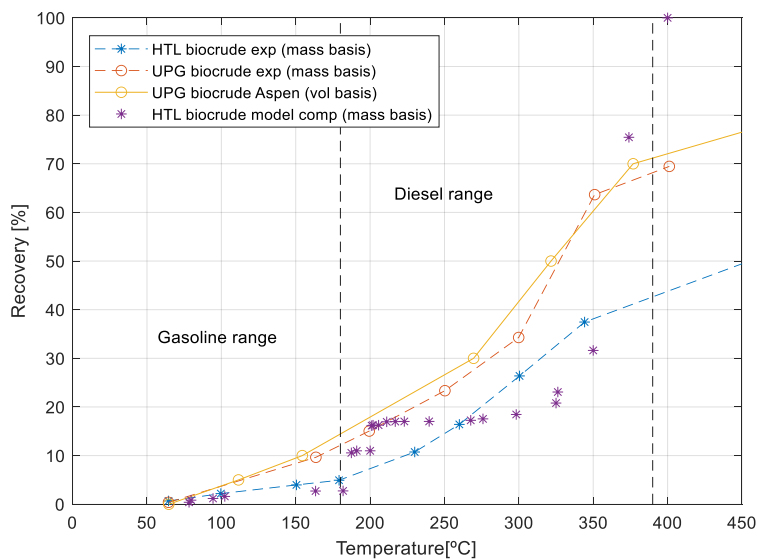

*Figure 5 True boiling point of HTL and upgraded biocrudes from simulation and experimental data* [15]

Regarding the upgraded biocrude, the properties are generated by the petro-characterization tools in Aspen Plus based on the experimental TBP curve supplied, and the small mismatch observed with the simulated is due to the different basis reported. The largest distillate fraction of about 60% is obtained in the diesel range, with lower recovery in the order of 10% in the gasoline range, as it has been previously reported for this type of feedstock [19,36]. Even though the TBP is accurately reproduced, there is a mismatch in some properties of Table 9 indicating that additional characterization data of the upgraded oil could be included for a more accurate representation.

### 3.2 Carbon dioxide capture and hydrogen purification-Dual Selexol™

The saturated HTL gas entering the absorption process has a $CO_2$ composition of 58 mol.% and is equivalent to 327 ton per day of $CO_2$ that can be potentially captured. The hydrotreater gas effluent is mainly hydrogen (98.8 mol%) with 0.1 mol% of $CO_2$, equivalent to additional 5.4 ton per day that are available for capture in the system for a total of 332.4 ton per day. Including both sources, the Selexol process yields a $CO_2$ recovery of 98 % and purity of 96.4 mol.%, equivalent to 325.7 ton per day. The hydrogen purity slightly increased to a final composition of 99%. From the $CO_2$ available, just 0.7 ton per day are lost during condensation of the HTL gas, other 4.5 ton per day are lost in the "combustible gas" stream, and 1.5 ton per day remain in the hydrogen rich stream. The results are discussed for all the steps: condensables removal from HTL gas, absorption towers and solvent regeneration. Finally, the sensitivity of the process over the HTL gas composition is presented. Regarding validation of the simulation results, due to the novelty of this application, there is not experimental data available on the carbon capture of the HTL gas with the DEPG solvent; however, there is published documentation by Aspen Tech® that validates the PC-SAFT property package used for acid gas cleaning with experimental and plant data at the conditions simulated. Please see reference [16] for more details of the validation.

#### 3.2.1 Removal of condensables from gas

The removal of condensables from the saturated HTL gas prior to absorption is achieved by cooling, and the $CO_2$ composition in the phases formed is shown in Figure 6. As the HTL gas is cooled down from the initial temperature (150 ºC), the concentration of $CO_2$ in the vapor increases due to the separation of condensables in the aqueous phase until a maximum of around 68% between 10-20 ºC. Below 20 ºC, the formation of a second liquid phase rich in $CO_2$ takes place causing a steeper decrease in the $CO_2$ available in the gas phase. Even though further cooling could be theoretically applied to directly obtain liquid $CO_2$ from the separator, the purity would still be below specification even at -20 ºC due to the condensation of other gas components. This evidences the difficulty in obtaining high purity liquid $CO_2$ by conventional V-L equilibrium despite of its relatively high concentration in the gas. Consequently, the temperature in the separator is set around 30 ºC in order to avoid $CO_2$ losses in the liquid, and the gas after separation is fed into the absorption column. In the hydrotreater gas there are not condensable compounds that can be separated by cooling, for which this step is not necessary.

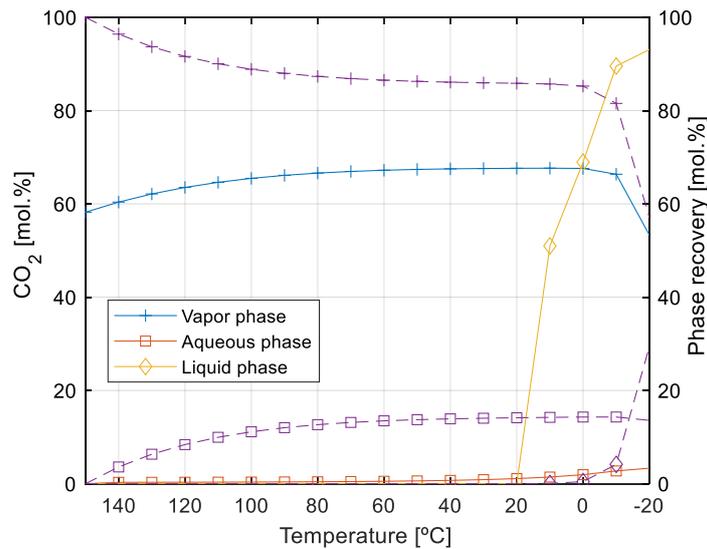

*Figure 6 Composition of $CO_2$ and recovery of the different phases after cooling of HTL gas at 40 bar*

### 3.2.2 Absorption and solvent regeneration

*Absorption columns:*

The change in composition of the HTL gas along the absorber can be observed in Figure 7a, being stage 8 the bottom of the column in which the gas is fed. As the gas flows upwards the concentration of $CO_2$ decreases while the concentration of hydrogen, methane and carbon monoxide increases due to their lower solubility in the solvent compared to $CO_2$ (Figure 7c). In the absorption of the hydrotreater gas (Figure 7b) the concentration of $CO_2$ and C3+ hydrocarbons also decreases from bottom to top, however, since these components are already in a very low concentration (less than 0.5 %), the impact of the absorption in the hydrogen and methane profiles is much lower. In this case, a higher pressure was used to increase the solubility and compensate for the lower ratio of solvent to gas compared to the first absorber.

From Figure 7c it can be seen that the solubility decreases with increasing temperature, reason for which at lower temperatures the demand of solvent is lower for a determined recovery (See also Figure 8). Even though the selectivity of $CO_2$ over propane slightly improves at higher temperatures, it decreases considerably with respect to ethane and the other compounds. The use of low temperatures in the solvent is in agreement with the reported in the consulted sources for the Selexol™ process [8,9,37].

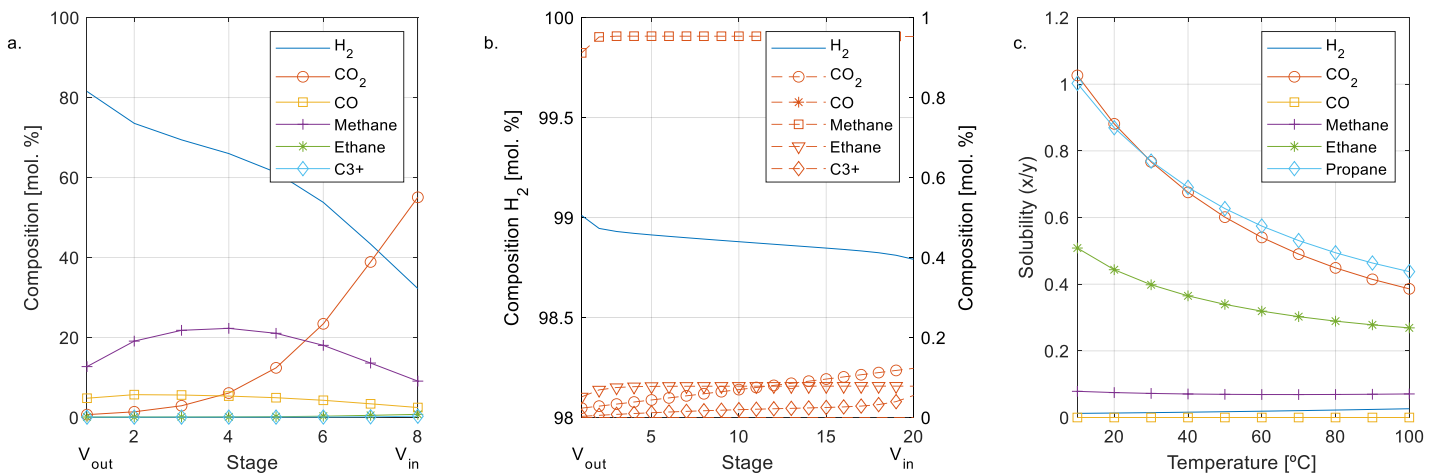

*Figure 7 Composition profile of vapor phase along the absorption column (a) HTL gas (b) Hydrotreater gas (c) Solubility of gas components at different temperatures and 40 bar*

The impact of the flow of solvent at 3 and 10 ºC in the purity and recovery of the $CO_2$ is presented in Figure 8a, expressed as liquid to vapor (L/V) ratio of the solvent relative to the fixed HTL gas inlet. For both temperatures the purity requirement is satisfied even at relatively low flows but with poor recoveries in the order of 40%. As the flow of solvent increases, purity and recovery increase until reaching a maximum after which the purity decreases due to a higher absorption of methane and ethane that are released during solvent recovery (Figure 8b). Furthermore, it is desirable to have lower solvent flows due to the energy consumption in the reboiler and pumps. Based on this the solvent flow is set corresponding to a ratio of 1.2, in order to maximize the recovery and meet purity while minimizing energy consumption.

*Solvent regeneration:*

After absorption of the HTL gas, the enriched solvent is subject to successive expansion and separation steps to release the $CO_2$ and other absorbed compounds. Overall, it is ideal to recover the less soluble gases from the solvent in the first step in order to facilitate the recovery of high purity $CO_2$ in subsequent expansions. In this case, two expansion stages were enough to obtain $CO_2$ of high purity. The purity and recovery of the $CO_2$ as function of the pressure level after the first expansion is shown in Figure 9. The purity requirement is not satisfied above 11 bar and can be explained as at higher pressure the solubility is higher and it is more difficult to release and recycle the less soluble impurities after the first expansion, being released in the second stage and affecting the purity of the $CO_2$. Therefore, the pressure is set at 9 bar and at 1 bar after the second expansion in order to guarantee maximum $CO_2$ recovery.

Table 10 shows the final configuration of the system and the composition of the produced streams is summarized in Figure 10a. The $CO_2$ is obtained at high purity and recovery, satisfying the minimum concentration of 95% taken as reference in this study. Regarding the hydrogen for recycle, a high recovery and purity was also obtained, having methane as the main impurity with a concentration of 0.9%. The purity level of the hydrogen is in the same order of the reported for commercial technologies such as membrane and pressure swing adsorption (PSA), typically used for hydrogen recovery in petroleum refining operations (purity between 93-99.9% and recovery in the range of 60-80% [38]). Regarding the methane, it is known that the impact in the performance of the hydrotreating depends on the type of catalyst and the threshold for commercial catalysts could be as low as 0.01- 0.05%. Such low concentrations would require a larger purge and higher excess hydrogen than the currently assumed, implying further adjustment in the Selexol™ process proposed, since at the conditions simulated the concentration slightly decreases to 0.7% after mixing with the make-up hydrogen. One possible modification is the addition of more solvent in the hydrogen absorber that can be regenerated directly in the reboiled column in a separated loop, improving the separation of methane without affecting the $CO_2$ absorption section, but increasing the energy demand for solvent regeneration.

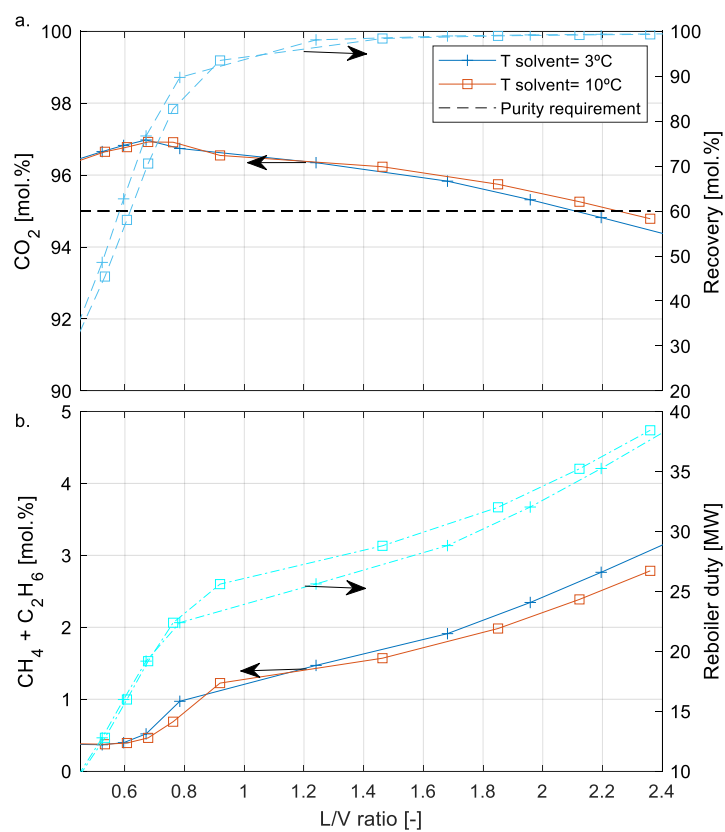

*Figure 8 Impact of flow ratio of solvent to HTL gas in (a) purity of $CO_2$ and recovery; (b) $C_1$-$C_2$ impurities and reboiler duty*

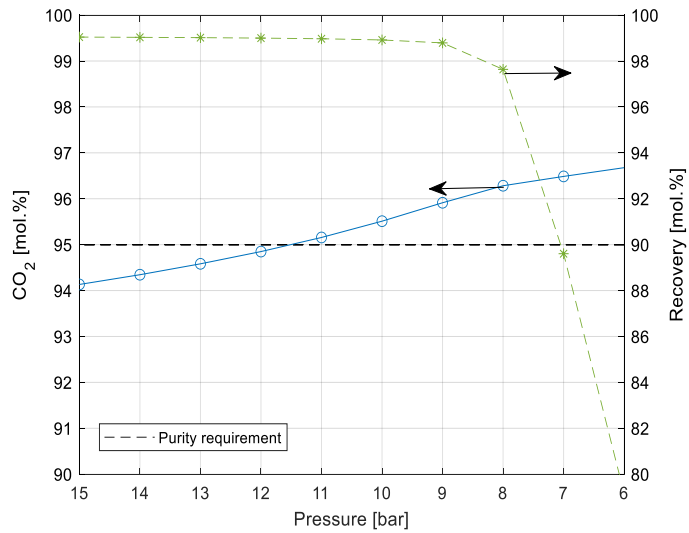

*Figure 9 Purity and recovery of $CO_2$ at different pressure levels during solvent regeneration*

*Table 10 Process design results for carbon dioxide capture from HTL gas using Selexol™ process*

| Absorption column | HTL gas | Hydrotreater gas |
|---|---:|---:|
| Pressure [bar] | 40 | 60 |
| Solvent temperature [ºC] | 7.0 | 3.0 |
| Number of separation stages | 8 | 20 |
| Solvent to vapor ratio (L/V) [-] | 1.2 | 0.3 |
| Solvent regeneration | | |
| Pressure levels [bar] | | 9; 1 |
| Number of stages reboiled column | | 10 |
| Solvent purity [mol.%] | | 99.9 |
| $CO_2$ recovery [%] | | 97.9 |
| $CO_2$ purity [mol. %] | | 96.3 |
| $H_2$ recovery [%] | | 95.6 |
| $H_2$ purity [mol. %] | | 99.0 |

### 3.2.3 Sensitivity of the process to initial HTL gas composition

The results of the sensitivity analysis of the process under different initial HTL gas composition are presented in Figure 10b, keeping other parameters in the simulation constant. It can be observed that the purity of the obtained $CO_2$ is very sensitive to changes in the initial composition of the HTL gas, being below the purity requirement in the cases where the initial $CO_2$ concentration is in the lower range and the C2+ hydrocarbons in the higher range reported in Figure 2. The purity increases with higher initial $CO_2$ concentrations but the recovery decreases if the flow of solvent is not increased accordingly. Furthermore, a larger increase in C2+ hydrocarbons seems to have a higher negative impact in the purity than the lower $CO_2$ concentration. This is explained by the similar or higher solubility reported for C2+ hydrocarbons in DEPG relative to the $CO_2$ that impedes their separation.

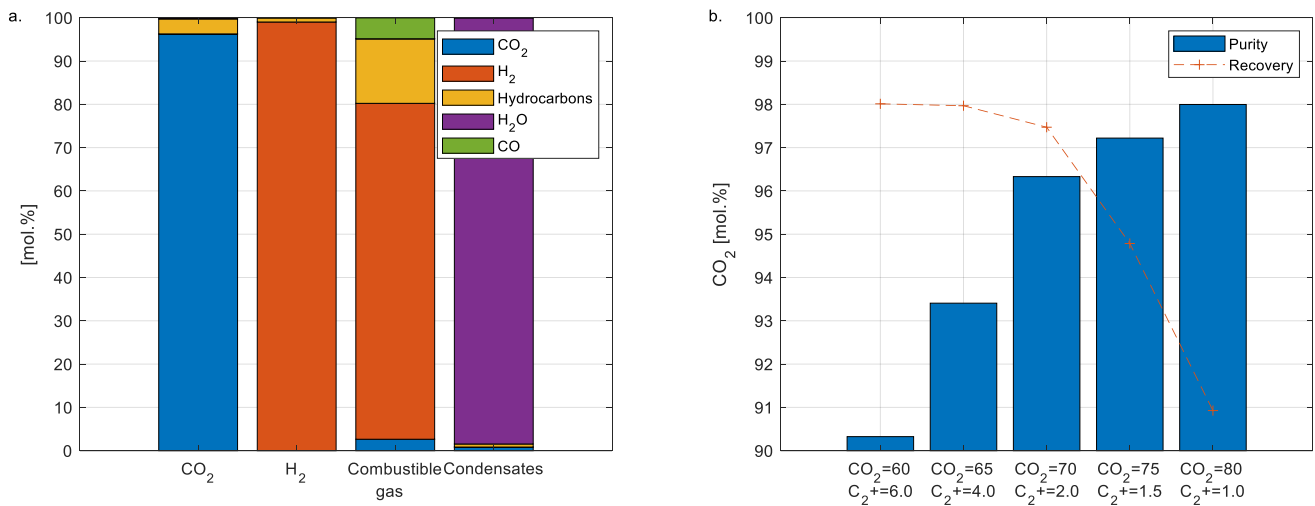

Figure 10 (a) Composition of products from Selexol™ process, (b) Sensitivy of purity and recovery of $CO_2$ to different HTL gas composition in mol. % at fixed solvent flow.

### 3.3 Heat integration and energy requirements

The hot and cold composite curves for both HTL base and Selexol™ sections are shown in Figure 11.

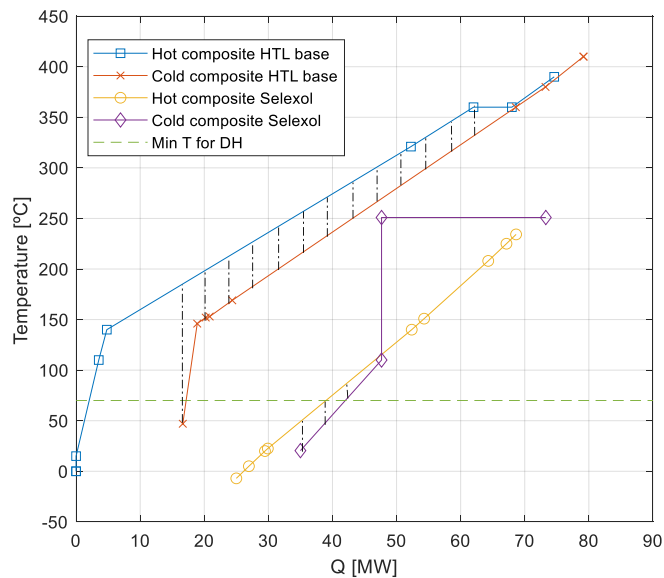

Figure 11 Hot and cold composite curves of HTL base and carbon capture sections (hot streams need to be cooled down and cold streams need to be heated)

The heat exchanged within process streams corresponds to the area between the hot and cold curves, and the segments outside must be covered by utilities. In the standard HTL base with upgrading, heat integration between hot and cold streams is possible and the pinch occurs at 390 ºC and in the Selexol™ section the pinch is at 100 ºC. Part of the heat demand can be supplied within the process but external heat is required for the HTL reactor (3.95 MW) and the reboiler column of the Selexol™ section (25.63 MW).

The heat produced in the process is estimated to 48 MW, and due to the relative high temperature, part of it can be considered for district heating use. The main source of excess heat comes from the Selexol™ process, as most of the heat available in the HTL base case is integrated within the process itself, being therefore almost self-sufficient in heat demand. Taking into account that the minimum temperature expected for 4[th] generation district heating systems is around 50-70 ºC, the cold utility is estimated as the heat below the green line (60 ºC + $\Delta T_{min}/2$) in Figure 11 (12 MW approximately) and the potential available is the remaining 36 MW.

The overall energy balance is shown in Figure 12 based on high heating values of the streams. The enthalpy values of all the streams can be consulted in the Supplementary material. The diagram shows the high efficiency of the HTL process towards the biocrude, which constitutes the main product of the process with the highest energy content, while the energy contained in the other streams hold the potential to be utilized within the process or in other purposes.

The excess heat is the main side-product in terms of energy, with potential to be used for district heating and also below 50 ºC in applications such as heat pumps for ultra low district heating (ULTDH). The heat integration is represented in the bottom of the diagram, showing that energy can be effectively integrated in the HTL and upgrading sections, and the external hot utility is mainly for the operation of the dual functionality in the Selexol™ unit, which can be potentially covered by the combustible gas produced.

Regarding the carbon dioxide, the energy content reflects the presence of hydrocarbon impurities and in the aqueous phase it corresponds to the presence of soluble organics that can be further treated through processes such as catalytic hydrothermal gasification (CHG), suggested as one of the alternatives for aqueous phase valorization in HTL processes, or anaerobic digestion [39]. This would increase the overall energy efficiency of the system. The error in the energy balance is around 3% and is expected due to the error reported in the elemental balance that has an impact on the heating values of the products used for the calculation of the energy balance.

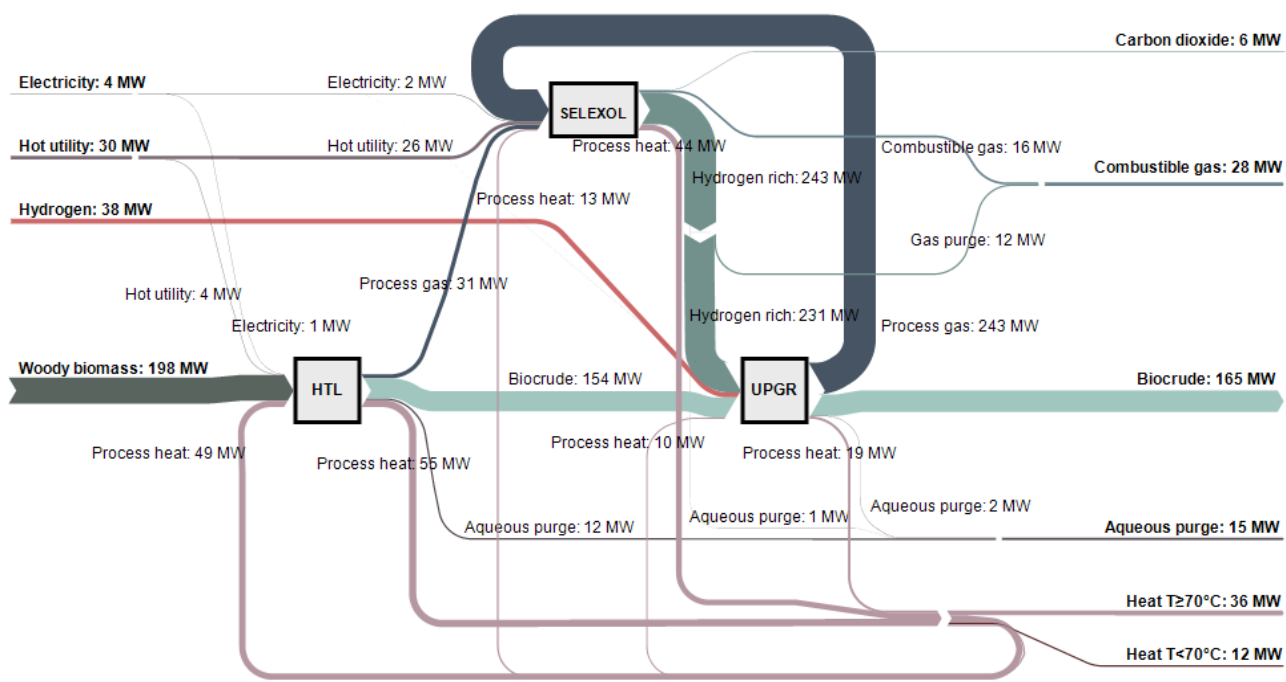

*Figure 12 Energy balance overall process. Tabulated data can be found in the supplementary material.*

### 3.4 Cost estimation

The results of the cost estimation are presented in Figure 13. In the base case, a minimum fuel selling price (MFSP) of 0.75 USD per liter of crude oil was obtained, equivalent to approximately 119 USD per barrel. The price of the hydrogen is the main contributor to the overall cost and for this analysis it is obtained via electrolysis (5.45 USD/kg in the base case). If a lower price of hydrogen was considered (1.90-5.12 USD/kg from SMR [6]), the estimated MFSP would be in the range of 0.46-0.69 USD/L, or 73-110 USD/barrel, which is in the range of the fossil crude around 70 USD per barrel in 2018 [40]. The revenues from the heat were estimated assuming that only 30, 40 and 50% of the process heat would be available in the minimum, base and maximum case respectively, as the precise value is not calculated in this study, and the combustible gas is not included as it is expected to be used for heating within the process. From Figure 13 it can be observed that the additional cost of the Selexol™ process in the base and minimum cases is almost covered by the potential revenue from the carbon dioxide emission allowances in the European market.

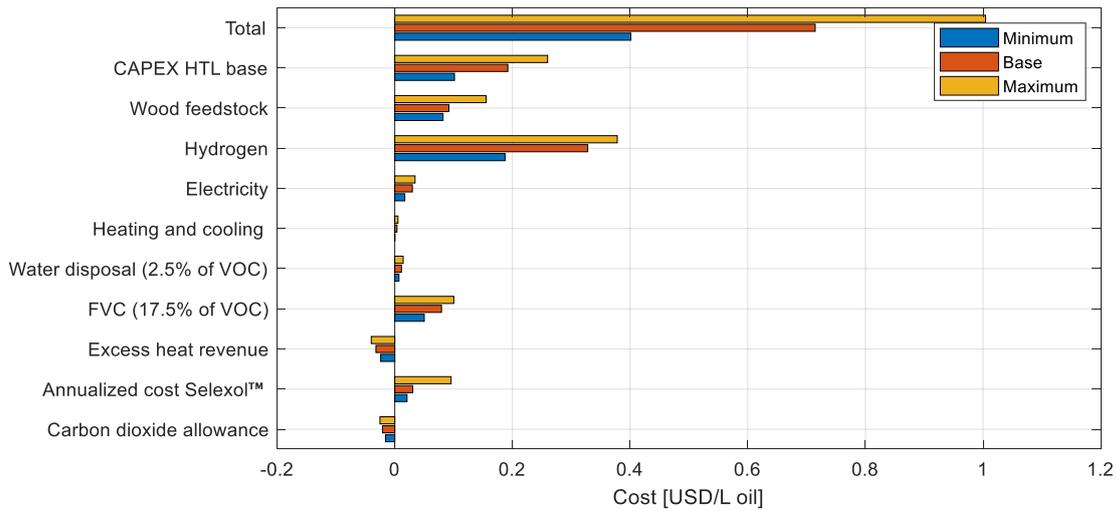

*Figure 13 Main contributors to minimum fuel selling price (MFSP)*

More detailed results for the Selexol™ section are summarized in Table 11. The capital cost estimated in this study is compared in Figure 14a with the reported for other carbon capture facilities of different capacity but with similar capture efficiency (85-90%) [41], indicating that the result obtained is in the order of magnitude expected. Even though the capture technology in the referenced cases is MEA, this absorption process requires similar type of equipment. The operational cost in Table 11 excludes the heating utility, as there is potential to cover this requirement by burning the combustible gas produced. The distribution of the installed cost in Figure 14b -excluding valves- shows that the compressors are the most expensive items with a share of 42% of the total. Lower capital and operational costs have been reported in literature of the Selexol™ process by the use of ejectors that recover the energy released during the expansion of high pressure streams and utilize it to simultaneously compress low pressure streams, achieving reductions in the capital cost up to 28% and up to 6% in operation costs [42]. Therefore, lower capital and operational costs could be obtained by adopting this strategy. The calculated cost of the avoided $CO_2$ is 75 USD/ton, in the range of the reported for BECCS [2].

*Table 11 Results of cost estimation for carbon capture section Selexol™*

|  | Estimated | Min | Max |
|---|---|---|---|
| CAPEX Selexol™ [US million] | 41.40 | -- | -- |
| OPEX Selexol™ [US million/y] | 3.96 | -- | -- |
| Avoided $CO_2$ [USD/ton] | 75.64 | 20.00 [2] | 175.00 [2] |
| Avoided $CO_2$ [USD/kg oil] | 0.08 | 0.02 | 0.18 |

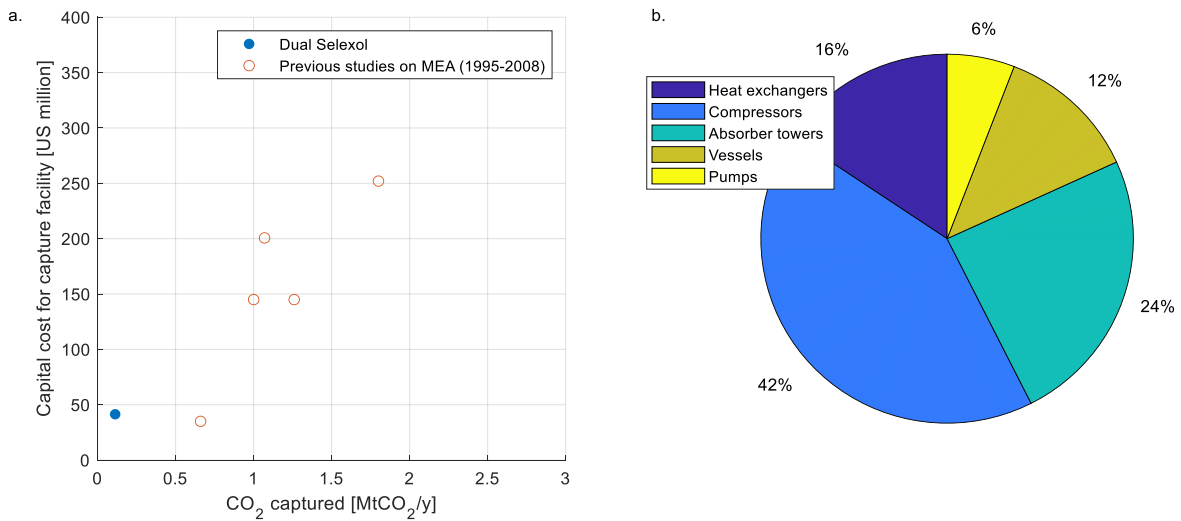

*Figure 14 (a) Capital cost of dual Selexol™ compared to the reported for MEA in [41] ;(b) Installed cost distribution estimated by Aspen Process Economic Analyzer.*

The impact on the MFSP of the HTL base case is summarized in Table 12 for a base, minimum and maximum case. In the base case the increase estimated is -0.3 %, indicating that there is not a significant net increase or decrease in the fuel price, so the additional cost of the Selexol™ process is covered by the revenues from the heat and carbon allowances. In case just the revenues from the heat are accounted the increase estimated is 3%. In the scenario of minimum price with highest revenues, the overall impact in the MFSP indicates that the revenues not only can compensate the additional cost of the Selexol™ but also decrease the biocrude cost by 15% and by 9% if the carbon allowance is not accounted. In the scenario of maximum price the increase estimated is 11%, however the cost of the Selexol™ is likely to be lower than the value assumed.

When the cost parameters are varied randomly, the results of the Monte Carlo simulations in Figure 15 show a similar scenario. When both revenues from heat and $CO_2$ are included, the MFSP curve is slightly shifted to the left indicating that is likely to have a decrease or similar cost than the base, varying from 6.97% cost reduction to 3.26% cost increase. When the $CO_2$ revenues are not included, the price is likely to be higher than the base, going from 2.38% cost reduction to 5.24% cost increase in agreement with the indicated previously.

Table 12 Impact of Selexol™, heat revenues and carbon allowances in the MFSP

|  | Base [USD/L] | Min [USD/L] | Max [USD/L] |
|---|---|---|---|
| Cost of biocrude in HTL base case | 0.74 | 0.45 | 0.95 |
| Cost of Selexol ™ | +0.07 | +0.02 | +0.15 |
| Revenues from heat | -0.05 | -0.06 | -0.04 |
| Revenues from carbon allowances | -0.02 | -0.03 | -0.02 |
| Total | 0.73 | 0.38 | 1.05 |
| Impact in biocrude cost [%] | -0.29 | -14.87 | 10.70 |

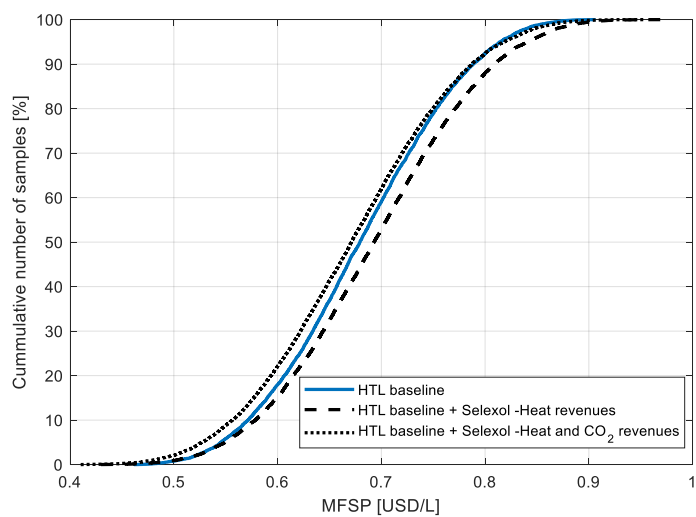

Figure 15 Impact of Selexol™, heat revenues and carbon allowances in the MFSP

### 3.5 GHG emission assessment

The results of the analysis are summarized in Figure 16. In the standard HTL base with upgrading, the GHG emissions were estimated in 19.4 kg $CO_{2eq}$/$GJ_{fuel}$ and represent a reduction of 85% compared to the fossil counterparts, estimated in 126 kg $CO_{2eq}$/$GJ_{fuel}$. The reduction is in the same range but slightly higher than the reported in [19] of 79% for the Canadian case, and can be attributed to the different consideration for the hydrogen production which in this case is via electrolysis and not SMR. Still, the contribution from the hydrogen supply is significant and depends on the nature of the electricity used for its production, which in this case is not 100% renewable as reflected in the electricity emission factor considered. Furthermore, since the dual Selexol™ is excluded in the HTL base, additional electricity for the hydrogen recovery step should be further accounted, however, is not expected to be significant. In a future scenario of utilizing 100% renewable electricity, the allocated GHG emissions would be close to 10 $CO_{2eq}$/$GJ_{fuel}$. This value is similar to the 13 $CO_{2eq}$/$GJ_{fuel}$ reported in a previous study with in-situ hydrogen supply for jet fuel production from HTL using forestry residues for a Swedish case [43]. Here the emissions allocated for raw material acquisition and transport were higher than the used in this study (2.5 vs. 6 kg $CO_{2eq}$/$GJ_{fuel}$), however, the emission reduction would only decrease from 85% to 82% if the same value was considered.

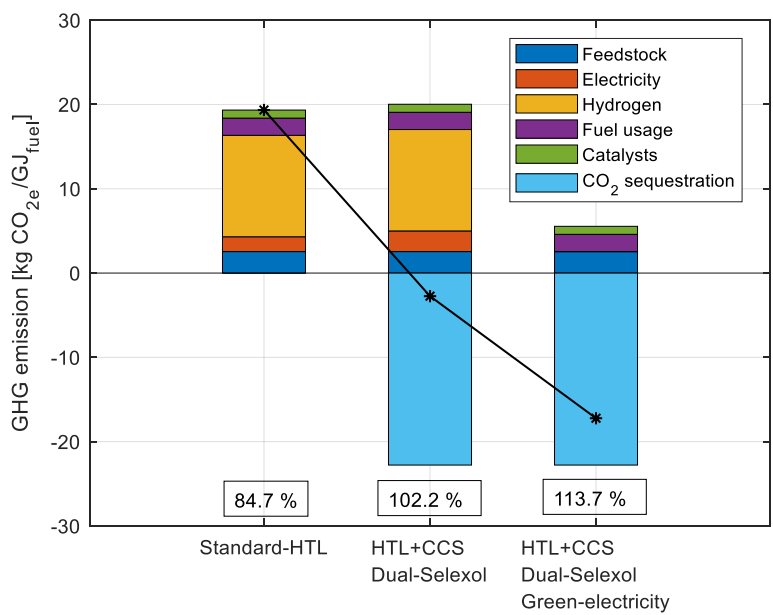

*Figure 16 GHG emissions and emission reduction percentage relative to fossil fuels*

The implementation of carbon capture in the process yields a net reduction in emissions from 19.4 to -2.7 kg $CO_{2eq}/GJ_{fuel}$, equivalent to 102% reduction compared to the fossil baseline, and further to -17.2 kg $CO_{2eq}/GJ_{fuel}$ if 100% renewable electricity is used, achieving 113.7% reduction. In this case, the different emissions allocated can be largely compensated by the effect of carbon sequestration, achieving carbon negativity in the process. In this analysis, the fractionation of the biocrude is not included and it is assumed that the heat requirement can be satisfied within the process due to the availability of the combustible gas. Even though these considerations should be further accounted for a more detailed estimation, these are not expected to drastically change the overall carbon negative picture in the scenario in which carbon capture is implemented in the process, given its significant impact in the emissions reduction.

The value of -17.2 kg $CO_{2eq}/GJ_{fuel}$ is higher than the reported by Cheng, Porter and Colosi [13], who recently estimated the global warming potential of HTT+CCS of lignocellulosic biomass in the order of -50 kg $CO_{2eq}/GJ_{fuel}$. The difference could be attributed to the different process configuration and the subcritical conditions used in this study that lead to lower energy consumption. The authors estimated the energy recovery of investment, EROI (EROI=Energy of product/Energy utilized in production process) in the order of 2, and identified a trade-off between energy yield and carbon sequestration. In this case, a higher EROI for the HTL+ Selexol™ has been estimated (Energy of biofuel and side products (combustible gas and excess heat) / Energy of utilities = 3.2) at expenses of lower $CO_2$ sequestration in agreement with this observation.

## 4. CONCLUSIONS

The process evaluated gives insight on different technical, economic and environmental aspects of the production of sustainable biocrude through HTL in combination with carbon capture via Selexol™. Overall, the results indicate that the Selexol™ process has potential for a cost-effective separation of the $CO_2$ in the HTL gas, delivering $CO_2$ at the required purity for pipeline transportation of 95 % with capture cost estimated in the order of 75 USD/ton.

Two main technical limitations were identified: The purity level showed high sensitivity to the initial gas composition, being below specification in cases where the concentration of $CO_2$ is in the lower range and the C2+ hydrocarbons in upper range of the reported for typical HTL gas. In terms of the dual functionality investigated for hydrogen purification, the methane concentration seems to be higher than the recommended for commercial catalysts in the upgrading step, therefore, other options commercially available may be further compared for hydrogen recovery in terms of technical performance and capital and operational costs.

For the overall HTL+carbon capture scheme, it is estimated that the revenues from the carbon emission market and the excess heat not only could compensate the additional cost of the carbon capture relative to the HTL baseline process, but also decrease the biocrude cost, having a relatively low impact in the MFSP. In case the revenues from the carbon allowances were not accounted due to the uncertainties in the carbon markets, the increase in the base case is estimated to be between 2.5 to 5% and expected to be below 10%. Even though this is moderated compared to CCS projects that report prohibitive capital expenditures, it is considered plausible due to the relatively high $CO_2$ concentration in the gas.

Additional costs of transportation and storage should be further considered, which are not included in this work and are related in a bigger picture to large-scale deployment of CCS infrastructure in Europe, regulatory framework and financial support among others.

The GHG emissions assessment indicates that the HTL + carbon capture has potential for a $CO_2$ reduction of 102% compared to the fossil baseline. It was observed that a higher reduction is hindered by the relatively large emissions allocated to the hydrogen production that in this case depend on the nature of the electricity used in the electrolysis process, and in case of 100% renewable electricity, it is estimated that there is potential for emission reductions up to 114%.

Overall, the results show that the technology has a promising performance as NET that can be of interest for future HTL projects in scenarios of high BECCS deployment for climate change mitigation, particularly in cases where high-energy yield alternatives are preferred. In the context of the energy transition, this is crucial for the sectors difficult to decarbonize such as long-haul transport, where the demand for sustainable advanced biofuels is expected to increase in the short and medium term.

## 5. ACKNOWLEDGMENTS

This project has received funding from the European Union's Horizon 2020 research and innovation program under grant no. 765515 (Marie Skłodowska-Curie ITN, ENSYSTRA) and grant no. 727531 (4refinery).

## 6. BIBLIOGRAPHY

[1] IEA Bioenergy. Bio-CCS and Bio-CCUS in Climate Change Mitigation. Task 41 Proj 5 2019. http://task41project5.ieabioenergy.com/ (accessed October 21, 2019).

[2] Fuss S, Lamb WF, Callaghan MW, Hilaire J, Creutzig F, Amann T, et al. Negative emissions—Part 2: Costs, potentials and side effects. Environ Res Lett 2018;13:063002. https://doi.org/10.1088/1748-9326/aabf9f.

[3] EU. Directive (EU) 2018/2001 of the European Parliament and of the Council of 11 December 2018 on the promotion of the use of energy from renewable sources. Off J Eur Union 2018;OJ L 328:82–209.

[4] Castello D, Pedersen T, Rosendahl L. Continuous Hydrothermal Liquefaction of Biomass: A Critical Review. Energies 2018;11:3165. https://doi.org/10.3390/en11113165.

[5] de Jong S, Hoefnagels R, Faaij A, Slade R, Mawhood R, Junginger M. The feasibility of short-term production strategies for renewable jet fuels – a comprehensive techno-economic comparison. Biofuels, Bioprod Biorefining 2015;9:778–800. https://doi.org/10.1002/bbb.1613.

[6] Pedersen TH, Hansen NH, Pérez OM, Cabezas DEV, Rosendahl LA. Renewable hydrocarbon fuels from hydrothermal liquefaction: A techno-economic analysis. Biofuels, Bioprod Biorefining 2018;12:213–23. https://doi.org/10.1002/bbb.1831.

[7] Searle SY, Malins CJ. Waste and residue availability for advanced biofuel production in EU Member States. Biomass and Bioenergy 2016;89:2–10. https://doi.org/10.1016/j.biombioe.2016.01.008.

[8] Padurean A, Cormos CC, Agachi PS. Pre-combustion carbon dioxide capture by gas-liquid absorption for Integrated Gasification Combined Cycle power plants. Int J Greenh Gas Control 2012;7:1–11. https://doi.org/10.1016/j.ijggc.2011.12.007.

[9] P. Field R, Brasington R. Baseline Flowsheet Model for IGCC with Carbon Capture. Ind & Eng Chem Res 2011;50:11306–12. https://doi.org/10.1021/ie200288u.

[10] Porter RTJ, Fairweather M, Kolster C, Mac Dowell N, Shah N, Woolley RM. Cost and performance of some carbon capture technology options for producing different quality CO2 product streams. Int J Greenh Gas Control 2017. https://doi.org/10.1016/j.ijggc.2016.11.020.

[11] Ohta T. Energy Carriers And Conversion Systems With Emphasis On Hydrogen - Volume I. Eolss Publishers; 2009.

[12] Peramanu S, Cox B., Pruden B. Economics of hydrogen recovery processes for the purification of hydroprocessor purge and off-gases. Int J Hydrogen Energy 1999;24:405–24. https://doi.org/10.1016/S0360-3199(98)00105-0.

[13] Cheng F, Porter MD, Colosi LM. Is hydrothermal treatment coupled with carbon capture and storage an energy-

producing negative emissions technology? Energy Convers Manag 2020;203:112252. https://doi.org/10.1016/j.enconman.2019.112252.

[14] Jensen CU, Rodriguez Guerrero JK, Karatzos S, Olofsson G, Iversen SB. Fundamentals of Hydrofaction[TM]: Renewable crude oil from woody biomass. Biomass Convers Biorefinery 2017;7:495–509. https://doi.org/10.1007/s13399-017-0248-8.

[15] Jensen CU. PIUS - Hydrofaction Platform with Integrated Upgrading Step. Aalborg University, 2018.

[16] Dyment J, Watanasiri S. Acid Gas Cleaning using DEPG Physical Solvents: Validation with Experimental and Plant Data. White Pap Aspen Tech 2015:1–18. https://www.aspentech.com/en/resources/white-papers/acid-gas-cleaning-using-depg-physical-solvents---validation-with-experimental-and-plant-data.

[17] Coker AK. Ludwig's applied process design for chemical and petrochemical plants. 4th ed. Boston: Elsevier Gulf Professional Pub.; 2007.

[18] Lozano EM, Pedersen TH, Rosendahl LA. Modeling of thermochemically liquefied biomass products and heat of formation for process energy assessment. Appl Energy 2019;254:113654. https://doi.org/10.1016/J.APENERGY.2019.113654.

[19] Jensen CU, Guerrero JKR, Karatzos S, Olofsson G, Iversen SB. Hydrofaction[TM] of forestry residues to drop-in renewable transportation fuels. Direct Thermochem. Liq. Energy Appl., Elsevier; 2018, p. 319–45. https://doi.org/10.1016/B978-0-08-101029-7.00009-6.

[20] Fahim MA, Alsahhaf TA, Elkilani A, Fahim MA, Alsahhaf TA, Elkilani A. Hydroconversion. Fundam Pet Refin 2010:153–98. https://doi.org/10.1016/B978-0-444-52785-1.00007-3.

[21] Rackley SA, Rackley SA. Absorption capture systems. Carbon Capture and Storage 2017:115–49. https://doi.org/10.1016/B978-0-12-812041-5.00006-4.

[22] Abbas Z, Mezher T, Abu-Zahra MRM. $CO_2$ purification. Part I: Purification requirement review and the selection of impurities deep removal technologies. Int J Greenh Gas Control 2013;16:324–34. https://doi.org/10.1016/J.IJGGC.2013.01.053.

[23] Jasiūnas L, Pedersen TH, Toor SS, Rosendahl LA. Biocrude production via supercritical hydrothermal co-liquefaction of spent mushroom compost and aspen wood sawdust. Renew Energy 2017;111:392–8. https://doi.org/10.1016/J.RENENE.2017.04.019.

[24] Pedersen TH, Grigoras IF, Hoffmann J, Toor SS, Daraban IM, Jensen CU, et al. Continuous hydrothermal co-liquefaction of aspen wood and glycerol with water phase recirculation. Appl Energy 2016;162:1034–41. https://doi.org/10.1016/J.APENERGY.2015.10.165.

[25] Hansen NH, Pedersen TH, Rosendahl LA. Techno-economic analysis of a novel hydrothermal liquefaction implementation with electrofuels for high carbon efficiency. Biofuels, Bioprod Biorefining 2019;13:660–72. https://doi.org/10.1002/bbb.1977.

[26] Oko E, Zacchello B, Wang M, Fethi A. Process analysis and economic evaluation of mixed aqueous ionic liquid and monoethanolamine (MEA) solvent for $CO_2$ capture from a coke oven plant. Greenh Gases Sci Technol 2018;8:686–700. https://doi.org/10.1002/ghg.1772.

[27] Mokhatab S. Handbook of natural gas transmission and processing : principles and practices / n.d.

[28] Speight JG, Speight JG. Gas cleaning processes. Nat Gas 2019:277–324. https://doi.org/10.1016/B978-0-12-809570-6.00008-4.

[29] Mulder, M., Perey, P. L. & Moraga JL. Outlook for a Dutch hydrogen market: economic conditions and scenarois. Centre for Energy Economics Research, University of Groningen; 2019.

[30] BMWi. Prices of electricity for the industry in Denmark from 2008 to 2018 (in euro cents per kilowatt hour) [Graph]. Statista 2019. https://www.statista.com/statistics/595800/electricity-industry-price-denmark/ (accessed November 14, 2019).

[31] Winther A, Wenzel H, Rasmussen K, Justesen S, Wormslev E, Porsgaard M. Nordic GTL – a pre-feasibility study on sustainable aviation fuel from biogas, hydrogen and $CO_2$. 2019.

[32] Markets Insider. $CO_2$ european emission allowances. Commodities 2019. https://origin.markets.businessinsider.com/commodities/co2-european-emission-allowances (accessed December 8, 2019).

[33]   ENERGINET. Energi Data Service 2020. https://www.energidataservice.dk/en/ (accessed February 18, 2020).

[34]   Nel electrolysers. Water electrolysers 2020. https://nelhydrogen.com/water-electrolysers-hydrogen-generators/ (accessed March 4, 2020).

[35]   Madsen RB, Bernberg RZK, Biller P, Becker J, Iversen BB, Glasius M. Hydrothermal co-liquefaction of biomasses – quantitative analysis of bio-crude and aqueous phase composition. Sustain Energy Fuels 2017;1:789–805. https://doi.org/10.1039/C7SE00104E.

[36]   Pedersen TH, Jensen CU, Sandström L, Rosendahl LA. Full characterization of compounds obtained from fractional distillation and upgrading of a HTL biocrude. Appl Energy 2017;202:408–19. https://doi.org/10.1016/J.APENERGY.2017.05.167.

[37]   Doctor RD, Molburg JC, Thimmapuram PR, Berry GF, Livengood CD. Gasification combined cycle: Carbon dioxide recovery, transport, and disposal. Argonne, IL: 1994. https://doi.org/10.2172/10190542.

[38]   Kohl AL, Nielsen RB, Kohl AL, Nielsen RB. Membrane Permeation Processes. Gas Purif 1997:1238–95. https://doi.org/10.1016/B978-088415220-0/50015-X.

[39]   Usman M, Chen H, Chen K, Ren S, Clark JH, Fan J, et al. Characterization and utilization of aqueous products from hydrothermal conversion of biomass for bio-oil and hydro-char production: a review. Green Chem 2019;21:1553–72. https://doi.org/10.1039/C8GC03957G.

[40]   EIA. Brent Crude oil prices from 2014 to 2020 (in U.S. dollars per barrel) [Graph]. Statista 2019. https://www.statista.com/statistics/409404/forecast-for-uk-brent-crude-oil-prices/ (accessed November 15, 2019).

[41]   Ho MT, Allinson GW, Wiley DE. Comparison of MEA capture cost for low $CO_2$ emissions sources in Australia. Int J Greenh Gas Control 2011. https://doi.org/10.1016/j.ijggc.2010.06.004.

[42]   Ashrafi O, Bashiri H, Esmaeili A, Sapoundjiev H, Navarri P. Ejector integration for the cost effective design of the Selexol$^{TM}$ process. Energy 2018;162:380–92. https://doi.org/10.1016/J.ENERGY.2018.08.053.

[43]   Tzanetis KF, Posada JA, Ramirez A. Analysis of biomass hydrothermal liquefaction and biocrude-oil upgrading for renewable jet fuel production: The impact of reaction conditions on production costs and GHG emissions performance. Renew Energy 2017;113:1388–98. https://doi.org/10.1016/J.RENENE.2017.06.104.